\documentstyle[prl,preprint,aps]{revtex}
\tightenlines

\begin{document}
\title{Quenching and Annealing in the Minority Game}
\author{E. Burgos\cite{byline} and Horacio Ceva\cite{byline2}}
\address{Departamento de F\'{\i}sica, Comisi\'on Nacional de
Energ\'{\i}a At\'omica,\\
Avda. del Libertador 8250, 1429 Buenos Aires, Argentina}
\author{R.P.J. Perazzo\cite{byline3}}
\address{Centro de Estudios Avanzados, Universidad de Buenos Aires,\\
Uriburu 950, 1114 Buenos Aires, Argentina}
\date{\today }
\maketitle

\begin{abstract}
We report the occurrence of quenching and annealing in a version of the
Minority Game (MG) in which the winning option is to join a given fraction
of the population that is a free, external parameter. We compare this to the
different dynamics of the Bar Attendance Model (BAM)where the updating of
the attendance strategy makes use of all the available information about the
system and quenching does not occur. We provide an annealing schedule by
which the quenched configuration of the MG reaches equilibrium and coincides
with the one obtained with the BAM.

\end{abstract}


\draft

In recent times considerable attention has been given to the description of
self-organization phenomena in multi agent models. These are geared to
explain how (and which) individual decisions give rise to cooperative
effects that show up in the behavior of the system as a whole. Such models
have applications in a variety of situations that range from routing of
messages in an information network to the way in which equilibrium is
(dynamically) achieved by an ensemble of economic agents.

A well known framework in this line of research is the Bar Attendance Model
(BAM) proposed by B Arthur in \cite{Arthur2}. In this model, $N$ agents have
to decide whether or not to go to a bar at a certain date. The customers
share the perception that the attendance should not exceed a value $n$, $n<N$
in order to avoid crowding. If the attendance is kept below that threshold,
they derive a positive utility. As all the agents accommodate their {\em
individual} attendance strategies seeking to increase their own utility, the
whole system self-organises keeping the attendance at the required value.

In recent times a simplified version of the BAM called the Minority Game
(MG) has received much attention \cite{Challet}-\cite{Johnson} .The main
simplification is that agents derive utility whenever their choice coincides
with that of the minority. This is equivalent to fix $n$ in the BAM at
$n=(N-1)/2$. The agents update their attendance strategies making use of
information concerning the attendance of the previous days.

Several variations of this model have been discussed in the literature.
They essentially differ in the way in which the adaptation process takes
place.  We restrict ourselves to consider the probabilistic version
\cite{Burgos} of the Johnson et al. model\cite{Johnson} with a slight
change in order to leave the accepted level of attendance $n$ as an
external free parameter.  Each agent has a given strategy expressed by a
probability $p_i$ ($i=1,2,\dots ,N$) to attend the bar.  If her option is
to go (not to go) and the attendance $A$ is $A\leq n$ $(A>n)$, the agent
gains a point.  Otherwise, the agent looses a point.  If her total number
of points falls below a fixed negative threshold her account is reset to
zero and her strategy $p_i$ is updated changing its value into one chosen
at random in the interval $(p_i-\delta p,p_i+\delta p)$.  Reflective
boundary conditions are used whenever necessary.  The parameter $\delta p$
is kept fixed.  If $n=(N-1)/2$ this framework coincides with the well
known MG.

The state of the many agent system is described by the density distribution
$P(p)$ that gives the fraction of agents that have a strategy within the
interval $(p,p+dp)$. Within the framework of the MG, an asymptotic
$P_{\infty}(p)$ is reached that displays a polarization of the population:
essentially half of the agents always go to the bar while the other half
never go thus keeping the average attendance close to $N/2$ as expected.

The evolution of the MG with $\mu =n/N\neq 1/2$ has been studied by Johnson
\cite{Johnson2} reporting that a situation can be reached in which no
further updating of the strategies takes place and the distribution $P_q(p)$
that is reached strongly differs from $P_\infty (p)$. In the present note we
investigate this effect in detail. We prove that this is a situation in
which the system is far from equilibrium. We also provide a prescription
that can be followed to effectively reach equilibrium starting from a
quenched configuration.

We have performed extensive numerical simulations starting with an initial
condition for $P(p)$ in which all agents have a strategy $p$ that is chosen
at random in the interval $(0,1)$ with uniform probability. We have used
several values of $N$ and changed $\mu $ above and below $1/2$. The overall
behavior of the many agent system has been found to be nearly symmetric
around $\mu =1/2$. Hence, we only show results for $\mu >1/2.$

We find that if $\mu \geq \mu _{cr}$ a quenched distribution $P_q(p)$ is
reached. We establish that quenching has occurred when no further updating
of strategies takes place. This naturally happens when all agents have more
than a sizable number of positive points. In what follows most of our
discussions assume that the prize for winning measured in units of the fine
for loosing is $G=1$. We explicitely remark when this is not the case.

Some examples of $P_q(p)$ are shown in Fig.\ref{distrip}.  It is worth to
remark three features.  First the value $P_q(p)=0$ for $0<p<1/2$ (or
$1/2<p<1$) , second that the width of the peak near $p=1/2$ is only a
function of $\delta p$ and third that the total attendance never reaches
the required fraction $\mu $. These features can qualitatively be
understood.  If one assumes that $\mu $ is large enough, agents face very
different situations depending whether they have $p<1/2$ or $p\simeq 1$.
In the first case the agents loose points frequently and therefore change
their strategies constantly increasing the value of $p$.  Those agents
with $p\geq 1/2$ will instead seldom loose and their strategies are rarely
updated.  The attendance will therefore not increase significantly from
the initial value and a quenched state is rapidly reached.

The above arguments indicate that the shape of $P_q(p)$ strongly depends
upon the initial conditions. This is the reason why we prefer to call this
effect ``quenching'' rather that ``freezing''. The latter suggests a new
ordering that bears no relationship with the initial conditions, while the
former indicates that a dynamics has been imposed that suddenly fixes the
internal parameters of the system at values that are close to those
corresponding to initial conditions and are far from equilibrium.

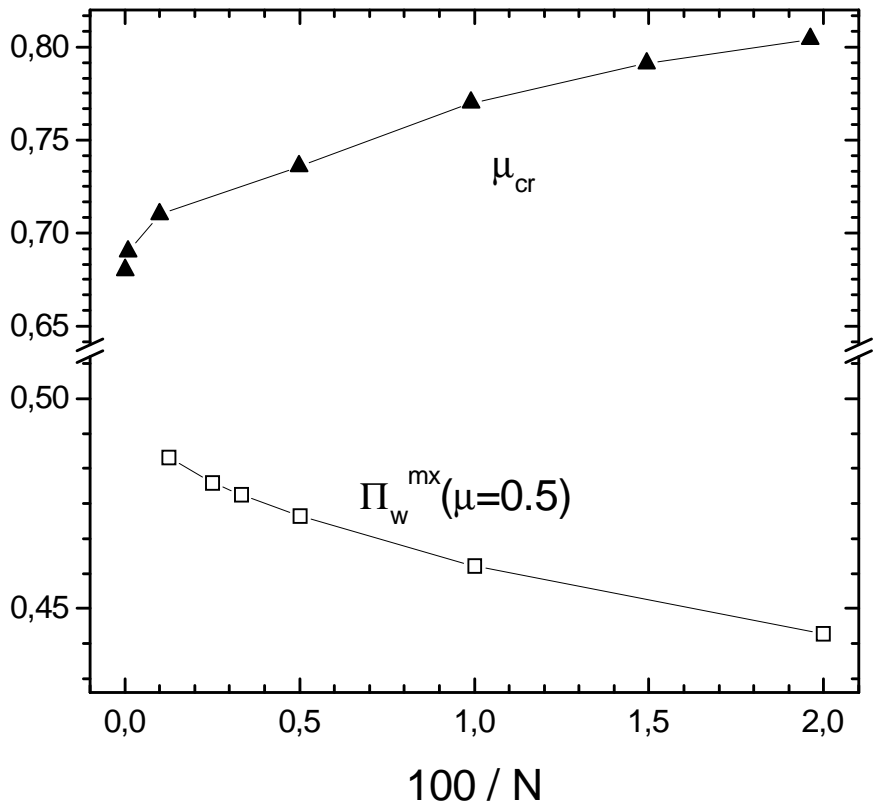
\begin{figure}[tbp]
\caption{{{{{{{\protect\small Plot of $\Pi^{mx}_w$ for $\mu=$1/2 as a
function of $100/N$. We also show $\mu_{cr}$ obtained for the initial
condition of a uniform distribution }}}}}}}
\label{finitescaling}
\end{figure}

Quenching starts to show up in a narrow region below a critical value $\mu
_{cr}$. The fact that this region becomes narrower for greater values of $N$
indicates that its existence is to be attributed to the fact that we are
dealing with a finite number of agents. For the same reasons the values $\mu
_{cr}$ also display some dependence upon $N$ (see Fig \ref{finitescaling}).
These values become smaller (larger) if $G$ is increased (diminished).

The occurrence of quenching can also be investigated in the BAM. According
to the rules of the BAM agents derive utility in two situations: one when
they decide to go to the bar and $A<n$ and when they decide not to attend
the bar and $A>n$. The adaptive dynamics of the BAM has been studied with
schemes either borrowed from genetic algorithms \cite{Arthur2} \cite
{Goldberg} \cite{nosotros} or bearing strong resemblances with that scheme.
A purely probabilistic, mean field approach is however desirable to ease the
comparison with the treatments of the MG.

We suggest a dynamics that is based upon the same information as in the MG
but uses it differently keeping track of the {\em reasons} for loosing or
winning. We introduce ``points'' and ``credits''. Points are gained or lost
if the option of {\em attending} the bar is correct or not. Credits are
instead gained or lost depending whether the option of {\em not attending}
the bar is correct or not. We therefore carry a double account: if an agent
chooses to attend (not to attend) the bar and $A<n$ ($A>n$)gains a point (a
credit). Otherwise, she looses a point (a credit). When either the number of
points or the number of credits falls below a negative threshold the
corresponding account is reset to zero and the strategy is changed in the
same fashion as already mentioned for the MG.

Using the above dynamics we have obtained the asymptotic density
distributions for various settings. We compare them with the ones obtained
with the MG in Fig.\ref{distrip}. Whenever $\mu <\mu_{cr}$ and no quenching
is found in the MG, both distributions $P_\infty ^{MG}$ and $P_\infty ^{BAM}$
can barely be distinguished from each other. In this regime while the
distribution remains stationary individual strategies are continously
updated and the level of attendance reaches the accepted fraction $\mu $.
However if $\mu >\mu _{cr}$ the distribution $P_\infty ^{BAM}(p)$ is
completely different to $P_q^{MG}(p)$ for the same settings. While in the MG
no further updating of strategies takes place, strategies in the BAM
continue to be updated at all times. In addition the density distribution
has lost all memory of the initial conditions and displays a strong
(asymmetric) self-organization: many agents attend the bar almost all the
time, while a smaller fraction have a very small probability to go.

It is possible to support these facts with a simple analytical discussion.
We note that if all agents share the same strategy $p$, the probabilities
$\Pi _{pt}$ and $\Pi _{cd\text{ }}$of winning respectively points and
credits are given by

\begin{eqnarray}
\Pi _{pt}(p) &=&S(N,n,p)=\sum_{i=0}^{n-1}{{N-1 \atopwithdelims() i}
}p^i(1-p)^{N-1-i}  \nonumber \\
\Pi _{cd}(p) &=&1-S(N,n+1,p)  \label{probapuntos}
\end{eqnarray}

for simplicity, we will write down the $N$ and $n$ dependence of $\Pi
_{pt,cd}$ only when necessary.  In order to be quenched the probabilities
$\Pi _{pt}$ and $\Pi _{cd}$ should both be greater than the probabilities
of loosing respectively points and credits, namely, $i.e.$ $\Pi _{pt}\geq
1/2,$ and $\Pi _{cd}\geq 1/2.$ This can clearly not be fulfilled
simultaneously.  Therefore agents either win points and loose credits or
vice versa forcing an updating of their strategies $p$.

A similar analysis can be made for the MG. If the system is prepared with a
distribution $P_0(p)=\delta (p-p_o)$ and $G=1$ this will be quenched if the
probability of winning $\Pi _w(p)$ fulfills $\Pi _w\geq 1-\Pi _w$, namely

\begin{equation}
\Pi _w(p_o)=p_o\Pi _{pt}(p_o)+(1-p_o)\Pi _{cd}(p_o) \geq \frac {1} {2}
\label{pganar}
\end{equation}

\noindent the case for $G\neq 1$ can easily be taken into account changing
the r.h.s. of Eq.\ref{pganar} into $1/(1+G)$.

The situation for the MG is completely different than for the BAM.  The
function $\Pi _w(p)$ has a single maximum in the interval $0\leq p\leq 1$.
Furthermore, the curves $\Pi _w(N,n,p)$ fulfill $\Pi _w(N,n,p)$ $\approx
\Pi _w(N,N-n,1-p)$.  Strict equality holds in the limit $N\rightarrow
\infty ,$ $\mu \neq 0,1.$ In Fig.\ref{sumaypganar} we show an example of
the functions $S(N,n,p)$ and $\Pi _w(p)$ for typical values of $N$ and
$n$, and $\mu >1/2$.

For finite $N$ and $\mu \simeq 1/2$ the value max$[\Pi _w]=\Pi _w^{mx}$ is
less than $1/2$ and no quenching occurs. For greater values of $\mu $ one
can define a critical value $\mu_{cr}=n_{cr}/N$ for which $\Pi_w^{mx}=1/2$
and quenching occurs. If one sets $\mu$= 1/2, the value of $\Pi_w^{mx}$ is
seen to approach the value 1/2 as $N\rightarrow \infty $ thus suggesting
that an infinite agent system prepared with the initial conditions
$P_0(p)=\delta (p-1/2)$ is almost quenched. (see Fig.
\ref{finitescaling}).

Above $\mu _{cr}$ Eq.\ref{pganar} is fulfilled for any $p_o$ in the interval
limited by $p_{<}$ and $p_{>}$ that are the two roots of the equation $\Pi
_w(p)=1/2$. Any system prepared with a distribution $P_0(p)=\delta (p-p_o)$
with $p_o$ within that region will rapidly be quenched $i.e.$ will not
change under the evolutive dynamics.

It can be checked that the system evolves quite differently if the initial
conditions are $p_o<p_{<}$ or $p_o>p_{>}$.  In the first case the whole
population of agents update their strategies, forcing all individual $p$'s
to surpass the value $p_{<}$.  Once this happens, updating stops and the
system gets quenched with an attendance level below the accepted fraction
$\mu $. For $p_o>p_{>}$ the asymptotic distribution $P_\infty ^{MG}$
displays a strong (asymmetric) polarization, updating of strategies
continues to take place continuously and the attendance reaches the
required fraction $\mu$.  We have checked numerically this with initial
conditions of the type $P_0(p)=\delta (p-p_o)$.  Typical examples of the
asymptotic distributions are shown in Fig.\ref{sumaypganar}.  Those that
correspond to initial conditions with $p_o>p_{>}$ are not distinguishable
from those obtained with BAM for the same settings.

The situations described hitherto indicate that outside a very restricted
region of $\mu$ and for different initial conditions, the dynamics of the MG
rapidly approaches quenching. The many agent system stops evolving, the
attendance remains fixed at a value that is far from the tolerated fraction
$\mu$ and all the other internal parameters remain close to the initial
conditions. The results reported in \cite{Johnson3} can be seen to fall also
in the framework presented here.

Opposed to this, the BAM approaches always a distribution
$P_{\infty}^{BAM}(p)$ that corresponds to what one expects of
thermodynamics
equilibrium: the attendance fluctuates around the expected fraction $\mu$,
all agents continue to update their individual strategies and
$P_{\infty}^{BAM}(p)$ remains stationary.

The way to regain equilibrium within the framework of the MG is through an
annealing protocol geared to correct progressively the quenched distribution
$P_q(p)$. The key point is to realize that quenching is produced by {\it the
memory stored in the points gained by the agents}. In fact the quenched
state remains fixed because there are no agents that ever loose points. We
have implemented a procedure that consists in obliterating periodically the
memory of the system by removing all the points gained by all the agents in
each annealing episode.

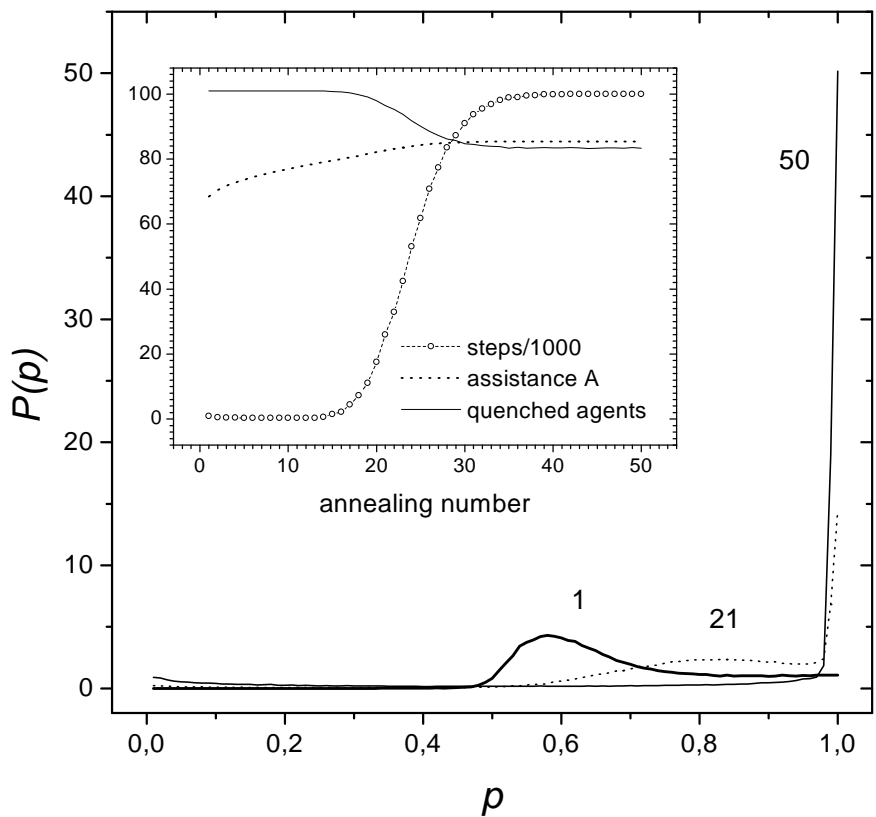
\begin{figure}[tbp]
\caption{{\protect\small Density distributions $P(p)$ obtained for the MG,
and $\mu=85/101$ after $a$ annealing episodes. The values of $a$ (1, 21,
50) are shown closed to the corresponding lines.The inital condition is a
uniform distribution. Each annealing episode involves a maximum of 100000
updating steps, after which all the points winned by the agents are removed.
In the inset we plot the variables that describe the whole process leading
to equilibrium.}}
\label{distriannealing}
\end{figure}

The results are displayed in Fig.\ref{distriannealing} for the case $\mu
=85/101$.  At the end of the first episode the distribution $P^1(p)$ is
entirely similar to those shown in Fig\ref{distrip}.  As agents are
repeatedly deprived of points some are increasingly forced to update their
strategies.  Successive changes approach $P_\infty ^{BAM}(p)$.  The way in
which the annealing protocol works is shown in the inset of Fig.\ref
{distriannealing}.  Quenching is checked by establishing that all agents
have more than, say, 5 points.  In addition we count how many steps are
required to reach such quenched state.  During the first 15 annealing
episodes very few steps are required in order to have all the 101 agents
with more than 5 points.  After that moment the number of (quenched)
agents that have more than 5 points starts to fall while the number of
steps grows rapidly thus showing the occurrence of a greater number of
individual updatings.  After 40 episodes full equilibrium is reached and
the number of steps saturates at the maximum allowed for each annealing
episode.  Throughout this process the attendance steadily grows
approaching the required fraction of $\mu \approx 0.85$.

In the present note we have discussed how quenching occurs in the MG. The
quenched state is far from equilibrium and strongly depends upon the initial
conditions. This situation is a direct consequence of the dynamics for
updating the individual attendance strategies. Contrary to this the BAM
reaches a situation that can be thought as thermodynamic equilibrium. This
difference has to be attributed to the fact the MG makes not full use of the
(global) available information about the system while the BAM dynamics keeps
instead track of the reasons for previous successes and failures. We also
show how the quenched state can be brought back to equilibrium through an
annealing protocol that amounts to periodically obliterate the memory stored
in the system by removing all the points that have been gained by the
agents. E.B. has been partially supported by CONICET of Argentina,
PICT-PMT0051;H.C. and R.P. were partially supported by EC grant
ARG/b7-3011/94/27, Contract 931005 AR.

\pagebreak

\begin{figure}[tbp]
\caption{Asymptotic distributions $P_q^{MG}(p)$ for $n= $ 70, 75, 80 and
$N=$101 agents.  Here $\mu_{cr}$ is 0.77.  Results for $\mu =$70/101 for
MG and BAM are essentially identical.  The BAM results for $\mu =$80/101
are shown with few scattered open squares.  These should be compared with
the dashed curve obtained for the MG with the same settings.}
\label{distrip}
\end{figure}

\begin{figure}[tcp]
\caption{The functions $S(N,n,p)$ and $\Pi_w(p)$ for $N=$101 and $\mu
=$75/101. The MG asymptotic distributions obtained with initial conditions
$P_o(p)=\delta (p-p_o)$ with $p_o=$0.4, 0.7 and 0.9 are referred to the
right axis. The results for $p_o=$0.9 for MG and BAM are essentially
identical The intersections of $\Pi _w(p)$ with the horizontal line at
1/2 are the two roots $p_{<}$ and $p_{>}$ }
\label{sumaypganar}
\end{figure}

\end{document}